\documentclass{webofc}
\usepackage[varg]{txfonts}   % Web of Conferences font
\usepackage{color}
\usepackage{graphicx}
\usepackage{multirow}
\usepackage{subfigure}
\usepackage{lineno}
\usepackage[rightcaption]{sidecap}

\newcommand{\bfg }{\begin{figure}[htpb]}
\newcommand{\efg }{\end{figure}}
\newcommand{\bmn }{\begin{minipage}}
\newcommand{\emn }{\end{minipage}}
\newcommand{\bt }{\begin{table}[htpb]}
\newcommand{\et }{\end{table}}

\newcommand{\sNN}{$\sqrt{s_{\rm NN}}$ }

\newcommand{\GeVc}{GeV/$c$}
\newcommand{\GeVcsq}{GeV/$c^2$ }

\newcommand{ \be }{\begin{equation}}
\newcommand{ \ee }{\end{equation}}
\newcommand{ \bea }{\begin{eqnarray}}
\newcommand{ \eea }{\end{eqnarray}}

\newcommand{ \pT}{$p_{\rm T}$}

\newcommand {\fz}	{f_{0}}
\newcommand {\ft}	{f_{2}}
\newcommand {\rhoz}	{\rho^{0}}
\newcommand {\nq}	{n_{q}}
\newcommand {\minv}	{m_{inv}}

\newcommand{\Rmnum}[1]{\expandafter\@slowromancap\romannumeral #1@}

\begin{document}
\title{NCQ scaling of $f_{0}(980)$ elliptic flow in 200 GeV Au+Au collisions by STAR and its constituent quark content}

\author{\firstname{Jie} \lastname{Zhao}\inst{1}\fnsep\thanks{\email{
			zhao656@purdue.edu
	}} (for the STAR collaboration)
}

\institute{Department of Physics and Astronomy, Purdue University
}

\abstract{
	Searching for exotic state particles and studying their properties have furthered our understanding of quantum chromodynamics (QCD).
    The $f_{0}$(980) resonance is an exotic state
    with relatively high
    production rate in relativistic heavy-ion collisions, decaying primarily into $\pi\pi$.
    Currently the structure and quark content of the $f_{0}$(980) are unknown
    with several predictions from theory being a $q\overline{q}$ state, a $qq\overline{q}\overline{q}$ state, a $K\overline{K}$ molecule state, or a gluonium state.
    We report the first $f_{0}$(980) elliptic flow ($v_{2}$) measurement from 200 GeV Au+Au collisions at STAR.
    The transverse momentum dependence of $v_{2}$ is examined and compared to those of other hadrons (baryons and mesons).
    The empirical number of constituent quark (NCQ) scaling is used to investigate the constituent quark content of $f_{0}$(980),
    which may potentially address an important question in QCD.
}
\maketitle
%\begin{linenumbers}

\section{Introduction}
Searching for exotic state particles and studying their properties have furthered our understanding of quantum chromodynamics (QCD).
Currently the structure and quark content of $f_{0}$(980) are unknown with several predictions being a $q\overline{q}$ state, a $qq\overline{q}\overline{q}$ state, a $K\overline{K}$ molecule state, or a gluonium state~\cite{Weinstein:1983gd,Weinstein:1990gu,Kleefeld:2001ds,Baru:2003qq,BESIII:2015rug,Achasov:2020aun}.
In contrast to the vector and tensor mesons, the identification of the scalar mesons is a long-standing puzzle~\cite{pdg:2020paz}. 
Previous preliminary experimental measurements~\cite{Fachini:2004jx} on the yield of $f_{0}$(980) at RHIC and theoretical calculation~\cite{ExHIC:2010gcb} suggest that it could be a $K\overline{K}$ stat.
In this analysis, the empirical number of constituent quark (NCQ) scaling~\cite{Molnar:2003ff,PHENIX:2003qra,STAR:2003wqp} is used to investigate the constituent quark content of $f_{0}$(980)~\cite{Gu:2019oyz}.

\section{Experiment setup and data analysis}
The data reported here are from Au+Au collisions at a nucleon-nucleon center-of-mass energy of 200 GeV, 
collected by the STAR experiment~\cite{Ackermann:2002ad} at Brookhaven National Laboratory in 2011, 2014 and 2016. 
A total of 2.4 billion minimum-bias (MB) events are selected for this analysis.
The main subsystem used for the data analysis 
is the Time Projection Chamber (TPC)~\cite{Anderson:2003ur} with 2$\pi$ azimuthal coverage at mid-rapidity. 
The TPC $dE/dx$ is used to select $\pi^{\pm}$ candidate with $0.2< p_{T} <5.0$ \GeVc. 

\begin{figure}[htbp!]
	\centering 
	\includegraphics[width=6.3cm]{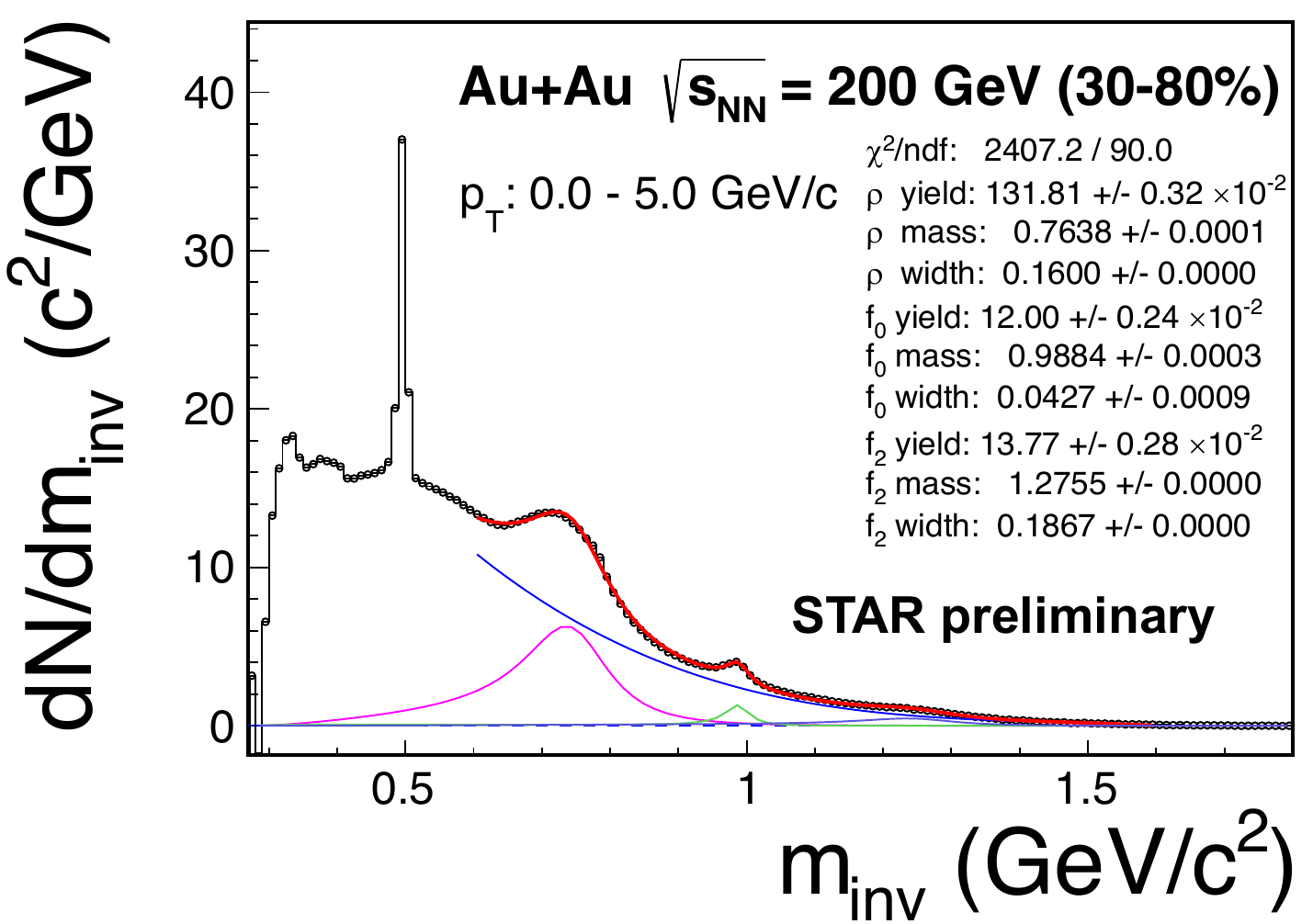} 
	\includegraphics[width=6.3cm]{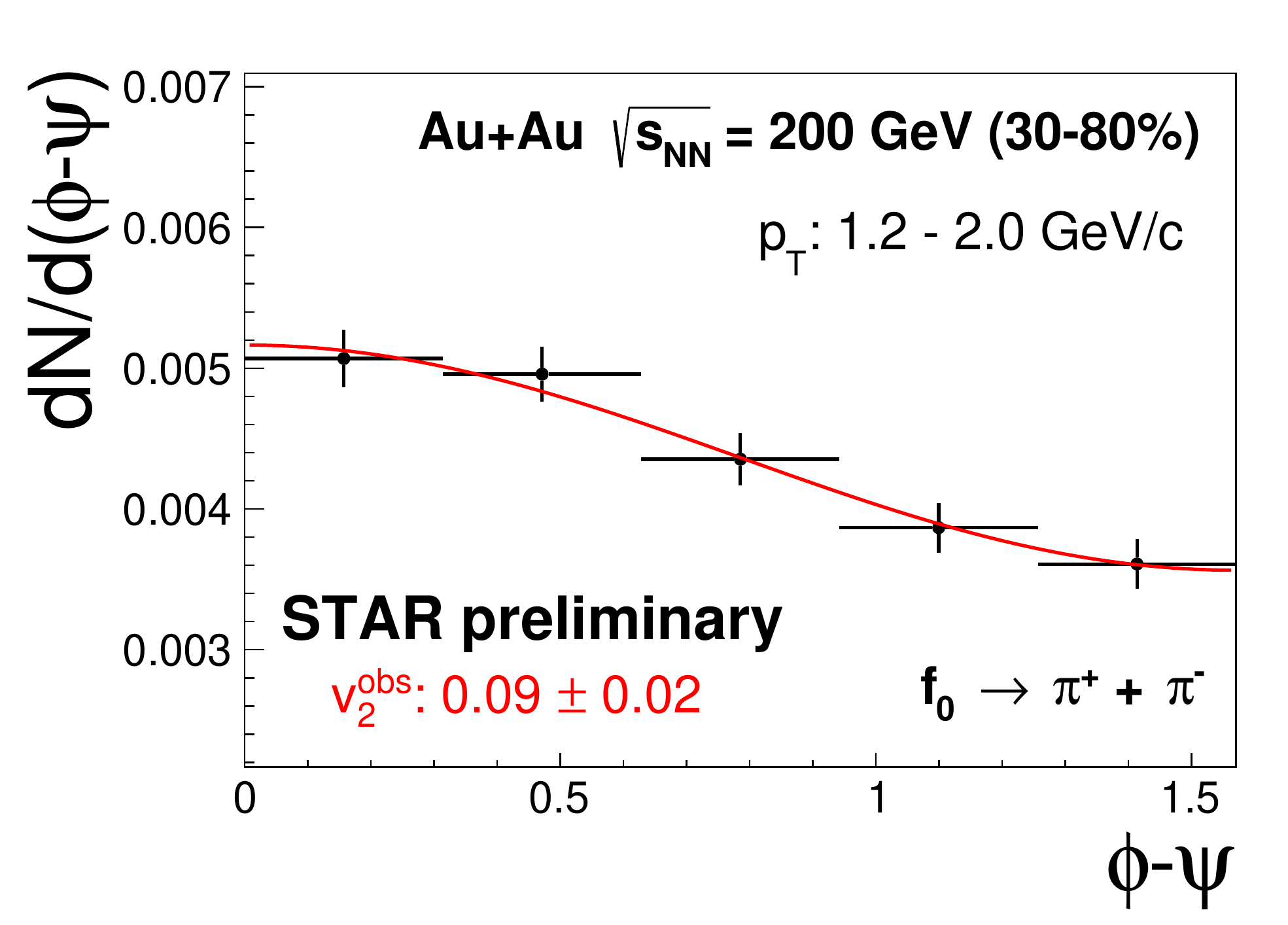} 
	\caption{(Color online)
	 (Left) The background subtracted $\pi^{+}\pi^{-}$ invariant mass distribution
	 over the $p_T$ range of $0<p_T<5.0$~GeV/$c$ in 30--80\% Au+Au collisions at $\sqrt{s_{NN}}$=200 GeV. 
	 The red line is the result of fit. 
	 The pink, green, violet lines represent the resonance peaks of the relativistic Breit-Wigner function. 
	 The solid blue line represents the residual background using a third order polynomial function. 
	 (Right) $f_{0}(980)$ yield as function of $\phi-\Psi$ in a given \pT\ bin.
	 Errors are statistical.
	 The red line represents a fit ($\propto(1+2v_{2}^{obs}\rm cos(2\phi-2\Psi))$) to the data. 
	}   
	\label{fig1}
\end{figure}

The $\pi^{+}\pi^{-}$ are used to reconstruct the $f_{0}(980)$. 
The combinatorial background subtraction is based on the mixed-event technique and the like-sign method~\cite{STAR:2015tnn}. 
The acceptance-corrected like-sign pairs~\cite{STAR:2015tnn,STAR:2013pwb} are used to subtract the combinatorial background 
after being normalized to unlike-sign pairs in the invariant mass ($\minv$) range beyond 1.5 \GeVcsq. 
Figure~\ref{fig1} (left) shows the background subtracted $\pi^{+}\pi^{-}$ invariant mass distribution. 
The resonance peaks are parametrized with the relativistic Breit-Wigner function~\cite{STAR:2002caw,STAR:2003vqj}.
The total fit function is given by:
%\begin{linenomath}
	\begin{equation}
		f(\minv)=\left(\sum_{X=\fz,\rhoz,\ft}\frac{A_X \minv m_X \Gamma(X)}{(\minv^2-m_X^2)^2 + m_X^2 \Gamma(X)^2}\right)\times PS + bg(\minv)
		\label{eqD}
	\end{equation}
%\end{linenomath}
where $\Gamma(X) = \frac{\Gamma_{X}m_{X}}{\minv}\left(\frac{\minv^2 - 4m_{\pi}^2}{m_{X}^{2}-4m_{\pi}^2} \right)^{J+1/2}$~\cite{STAR:2002caw,STAR:2003vqj},
$PS = \frac{\minv}{\sqrt{\minv^2+p_{T}^2}}\exp\left(-\frac{\sqrt{\minv^2+p_{T}^2}}{T}\right)$ 
is the phase space correction taking into account the $\pi\pi$ scattering during the hadronic phase~\cite{Shuryak:2002kd,Kolb:2003bk,Rapp:2003ar,STAR:2003vqj},
and $bg(\minv)$ is a third order polynomial function to describe the residual background.
$m_{X}$ and $\Gamma_{X}$ are the mass and width of the corresponding resonances. 
$\Gamma_{\rhoz}$ is set to 160 MeV, and $m_{\ft}$ and $\Gamma_{\ft}$ are set according to the PDG values~\cite{pdg:2020paz}. 
T is the kinetic freeze-out temperature, set to 120 MeV~\cite{Shuryak:2002kd}.
$A_{\fz}$, $A_{\rhoz}$, $A_{\ft}$, $m_{f_0}$, $\Gamma_{f_0}$, and $m_{\rho^{0}}$ are free parameters.

The event-plane method~\cite{Poskanzer:1998yz} is used to study the elliptic flow ($v_{2}$) of $f_{0}(980)$. 
The event-plane is reconstructed by all charged particles in the TPC with pseudorapidity $|\eta| < 1$ and transverse
momentum $0.2< p_{T} <5.0$ \GeVc. 
For each $\pi\pi$ pair, the two $\pi$ candidates are removed from the event-plane reconstruction to avoid auto-correlation.
The event-plane resolution is calculated by the correlation between two randomly divided sub-events from the full TPC~\cite{Poskanzer:1998yz}. 
Wide centrality bin effect is corrected by weighting the event-plane resolution with the $f_{0}(980)$ yield in each narrow centrality bin of 10\% size~\cite{STAR:2015gge}. 
Figure~\ref{fig1}(right) shows the $f_{0}(980)$ yield as function of the azimuthal angle difference between the $\pi\pi$ pair ($\phi$) and the event-plane direction ($\Psi$) in an example \pT\ bin.  

\begin{SCfigure}
    \centering
	\includegraphics[width=6.4cm]{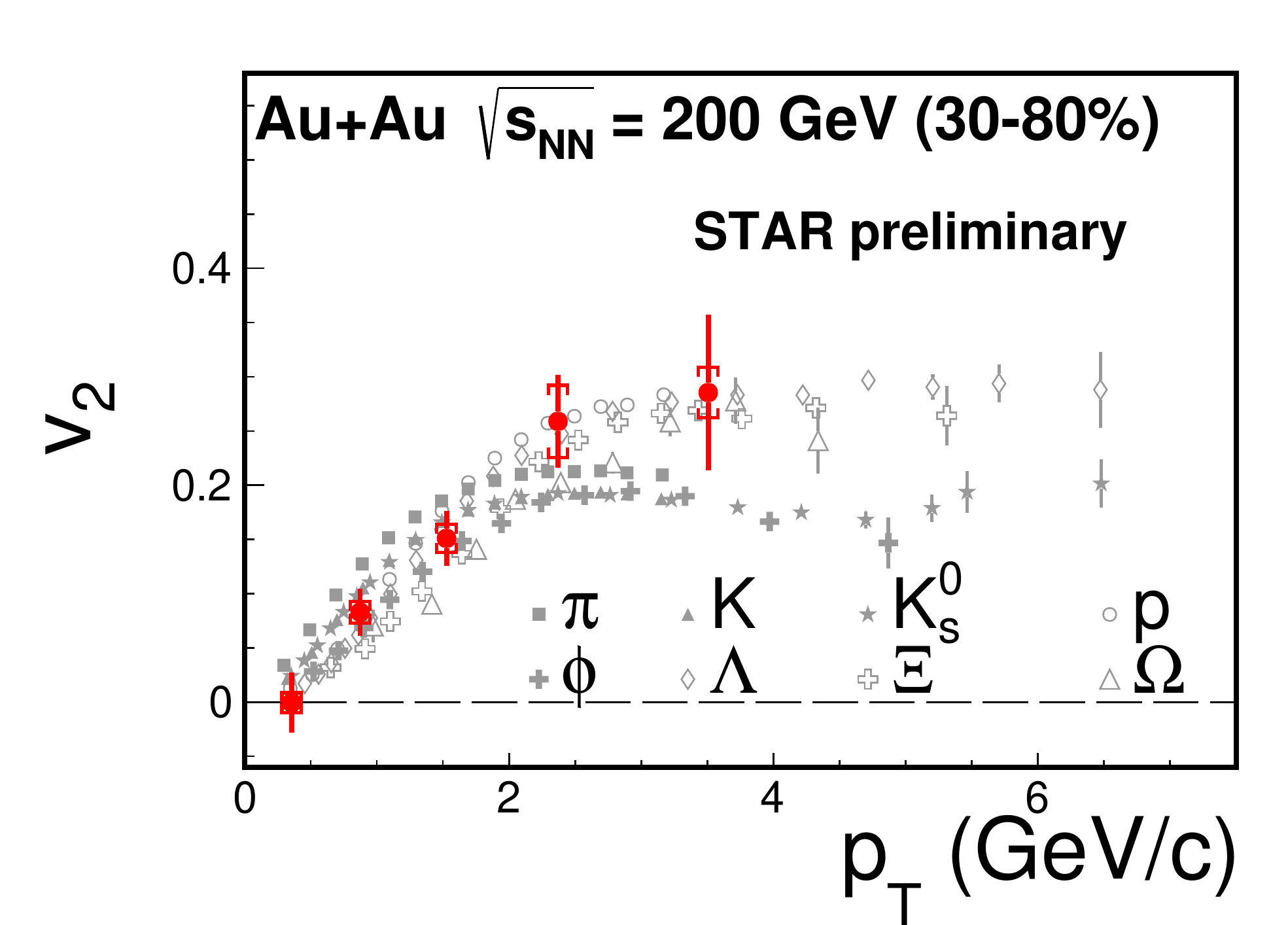}
    \caption{(Color online)
	  $f_{0}(980)$ $v_2$ as a function of \pT\ in 30-80\% centrality Au+Au collisions at $\sqrt{s_{NN}}$ = 200 GeV.	
	  Statistical uncertainties are shown by the vertical bars and systematic uncertainties are shown by the caps.
	  Results of other particles are taken from Ref.~\cite{STAR:2015gge}.
    }
    \label{fig2}
\end{SCfigure}

%\section{Results}
Figure~\ref{fig2} shows $f_{0}(980)$ $v_2$ as a function of \pT\ in 30-80\% centrality Au+Au collisions at $\sqrt{s_{NN}}$ = 200 GeV.
Results are compared with other identified particles: $\pi, K, p, K_{s}^{0}, \Lambda, \Xi, \Omega, \phi$~\cite{STAR:2015gge}.
In the low \pT\ region, the $f_{0}(980)$ $v_2$ seems to follow the mass ordering. 
In the higher \pT\ region, the $f_{0}(980)$ $v_2$ seems closer to the baryon band.

\begin{SCfigure}
	\centering 
	\includegraphics[width=8.7cm]{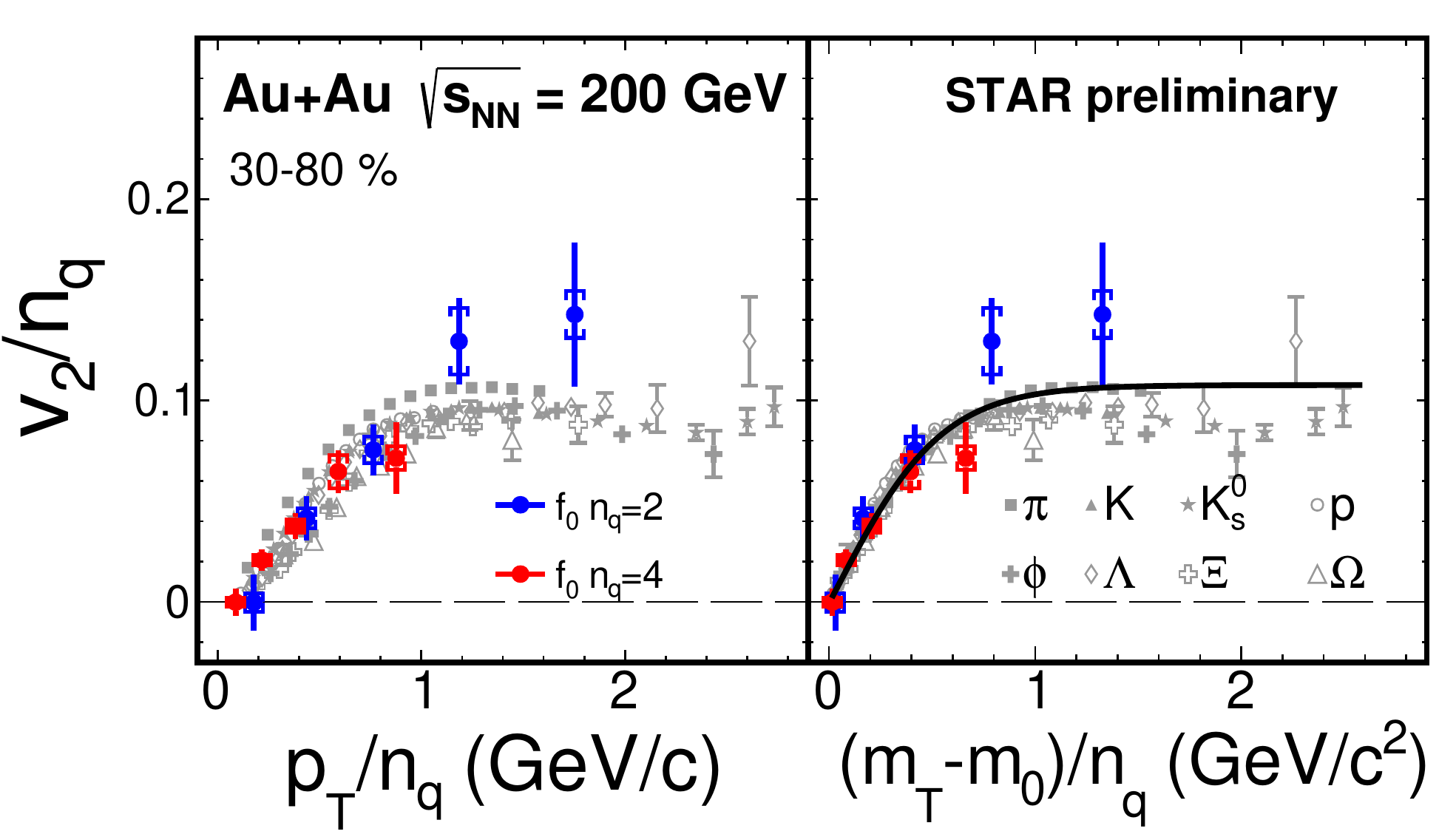} 
	\caption{(Color online)
		$f_{0}(980)$ $v_2$ divided by $\nq$ as a function of \pT$/\nq$ (left) and $(m_T-m_0)/\nq$ (right) in 30-80\% centrality Au+Au collisions at $\sqrt{s_{NN}}$ = 200 GeV.	
		Results of other particles are taken from Ref.~\cite{STAR:2015gge}.
		Black line in the right panel represents a fit 
		to results of other particles using a NCQ scaling inspired function (Eq.~\ref{eqnq}).
	}   
	\label{fig3}
\end{SCfigure}

Figure~\ref{fig3} shows the number of constituent quark ($n_{q}$) scaled $v_2$ as function of the $n_{q}$ scaled \pT\ (left) and $m_T-m_0$ (right). 
Here the $f_{0}(980)$ is assumed to have either 2 quarks or 4 quarks. 
The data are compared to the fit of other particles~\cite{STAR:2015gge} using a NCQ scaling inspired function~\cite{Dong:2004ve}: 
%\begin{linenomath}
\begin{equation}
	f_{v_2}(n_q)=\frac{an_{q}}{1+\exp(-((m_T-m_0)/n_q-b)/c)}-dn_q. 
	\label{eqnq}
\end{equation}
%\end{linenomath}
The 2-quarks (4-quarks) scaled $f_0(980)$ $v_2$ seems to deviate from the fit, above (below) the fit by $\sim1\sigma$ for the last one or two points at high $(m_T-m_0)/\nq$.  

\begin{SCfigure}
    \centering
	\includegraphics[width=6.4cm]{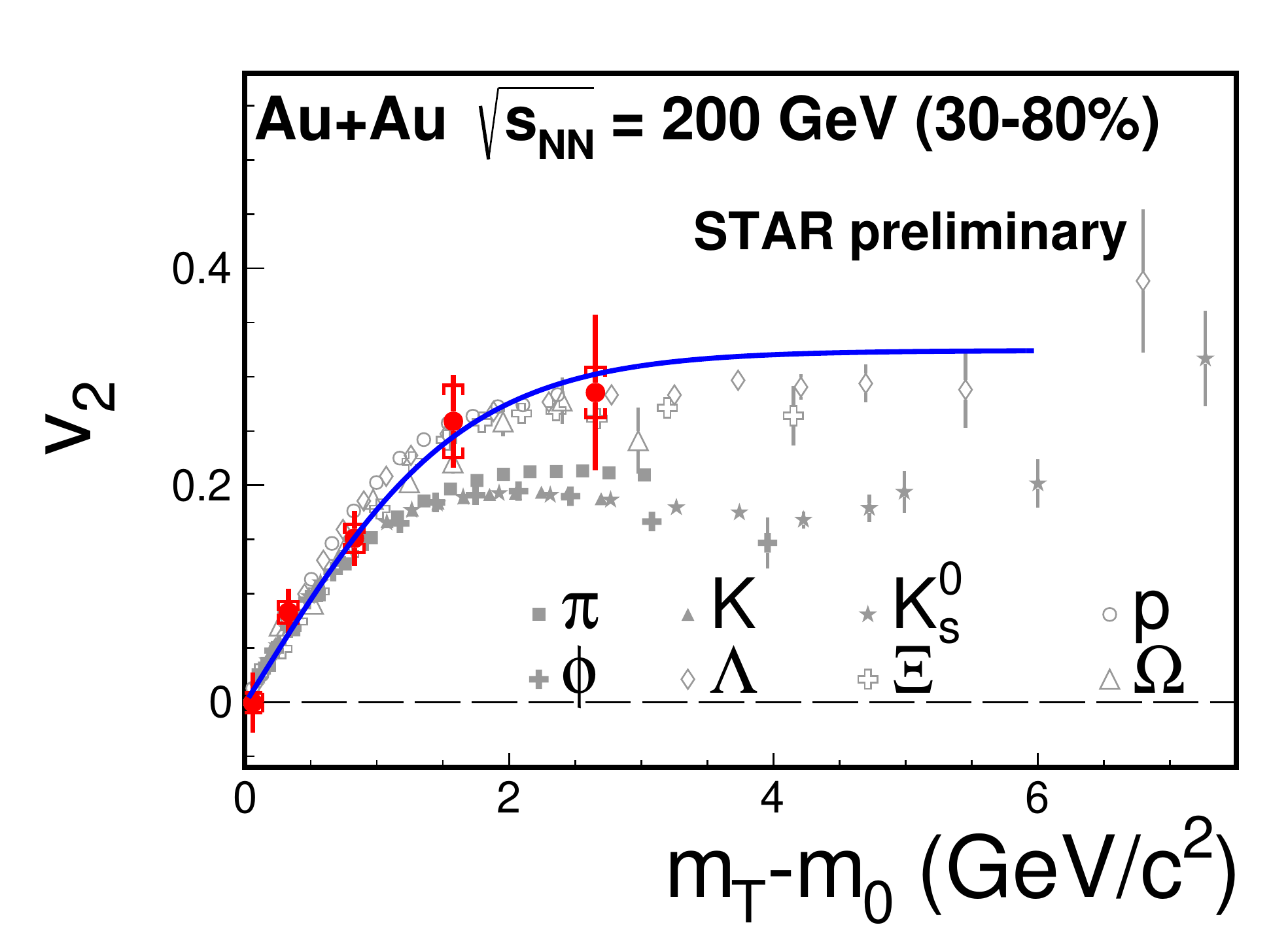}
    \caption{(Color online)
	  $f_{0}(980)$ $v_2$ as a function of $m_T-m_0$ in 30-80\% centrality Au+Au collisions at $\sqrt{s_{NN}}$ = 200 GeV.	
	  The blue curve represents the NCQ inspired fit (Eq.~\ref{eqnq}), 
	  where the only free parameter is the $\nq$ of $f_0(980)$
	  and all other parameters are fixed according to the fit in the right panel of the Fig.~\ref{fig3}.
    }
    \label{fig4}
\end{SCfigure}

Figure~\ref{fig4} shows $f_0(980)$ 
$v_2$ as a function of $m_T-m_0$ with a fit 
according to the function shown in Eq.~\ref{eqnq}.
In the fit, only the $\nq$ of $f_0(980)$ is treated as a free parameter and all other parameters are fixed according to the fit in the right panel of the Fig.~\ref{fig3}.
This NCQ scaling fit of the $f_0(980)$ $v_2$ yields $\nq = 3.0\pm0.7\mbox{ (stat)}\pm0.5\mbox{ (syst)}$. 

%\begin{figure}[htbp!]
%\end{figure}

With the current uncertainty, our result is not able to determine whether $f_{0}(980)$ is
a $q\overline{q}$, $qq\overline{q}\overline{q}$, $K\overline{K}$ molecule, gluonium state, or produced through $\pi\pi$ coalescence. 
It could also be given by some combined states as well.
Future measurements, e.g. the $f_{0}(980)$ yields, could also provide different aspect to understand it.

\section{Summary}
Preliminary results on the $f_0(980)$ $v_2$ in 30-80\% centrality Au+Au collisions at  
\sNN = 200 GeV are presented.
In the low \pT\ region (\pT<2 \GeVc), the $f_{0}(980)$ $v_2$ seems to follow the mass ordering.
In the higher \pT\ region (\pT>2 \GeVc), the $f_{0}(980)$ $v_2$ seems closer to the baryon band.
A NCQ scaling inspired function was used to fit the $f_0(980)$ $v_2$. 
The extracted quark content of $f_{0}(980)$ is $\nq = 3.0\pm0.7\mbox{ (stat)}\pm0.5\mbox{ (syst)}$. 
More data are needed to understand whether $f_{0}(980)$ is 
a $q\overline{q}$, $qq\overline{q}\overline{q}$, $K\overline{K}$ molecule, gluonium state, 
or produced through $\pi\pi$ coalescence. 
Our study indicates that heavy-ion collisions can be a useful place to examine the quark content of scalar mesons.
The isobar data taken in 2018 at RHIC and the 8-fold increase in Au+Au data
expected in 2023-2025 would provide more insights.

$\bold{Acknowledgments}$ 
This work was partly supported by the U.S. Department of Energy (Grant No. de-sc0012910).
%\end{linenumbers}   

\bibliographystyle{unsrt}
\bibliography{./ref}

\begin{thebibliography}{25}

\bibitem{Weinstein:1983gd}
J.D. Weinstein, N.~Isgur, Phys. Rev. D \textbf{27}, 588 (1983)

\bibitem{Weinstein:1990gu}
J.D. Weinstein, N.~Isgur, Phys. Rev. D \textbf{41}, 2236 (1990)

\bibitem{Kleefeld:2001ds}
F.~Kleefeld, E.~van Beveren, G.~Rupp, M.D. Scadron, Phys. Rev. D \textbf{66},
  034007 (2002)

\bibitem{Baru:2003qq}
V.~Baru et~al., Phys. Lett. B \textbf{586}, 53 (2004)

\bibitem{BESIII:2015rug}
M.~Ablikim et~al. (BESIII), Phys. Rev. D \textbf{92}, 052003 (2015)

\bibitem{Achasov:2020aun}
N.N. Achasov et~al., Phys. Rev. D \textbf{103}, 014010 (2021)

\bibitem{pdg:2020paz}
P.A. Zyla et~al. (Particle Data Group), Prog. Theor. Exp. Phys. \textbf{083C01}
  (2020)

\bibitem{Fachini:2004jx}
P.~Fachini, J. Phys. G \textbf{30}, S735 (2004)

\bibitem{ExHIC:2010gcb}
S.~Cho et~al. (ExHIC), Phys. Rev. Lett. \textbf{106}, 212001 (2011)

\bibitem{Molnar:2003ff}
D.~Molnar, S.A. Voloshin, Phys. Rev. Lett. \textbf{91}, 092301 (2003)

\bibitem{PHENIX:2003qra}
S.S. Adler et~al. (PHENIX), Phys. Rev. Lett. \textbf{91}, 182301 (2003)

\bibitem{STAR:2003wqp}
J.~Adams et~al. (STAR), Phys. Rev. Lett. \textbf{92}, 052302 (2004)

\bibitem{Gu:2019oyz}
A.~Gu, T.~Edmonds, J.~Zhao, F.~Wang, Phys. Rev. C \textbf{101}, 024908 (2020)

\bibitem{Ackermann:2002ad}
K.H. Ackermann et~al. (STAR), Nucl. Instrum. Meth. \textbf{A499}, 624 (2003)

\bibitem{Anderson:2003ur}
M.~Anderson et~al., Nucl. Instrum. Meth. \textbf{A499}, 659 (2003)

\bibitem{STAR:2015tnn}
L.~Adamczyk et~al. (STAR), Phys. Rev. C \textbf{92}, 024912 (2015)

\bibitem{STAR:2013pwb}
L.~Adamczyk et~al. (STAR), Phys. Rev. Lett. \textbf{113}, 022301 (2014)

\bibitem{STAR:2002caw}
C.~Adler et~al. (STAR), Phys. Rev. Lett. \textbf{89}, 272302 (2002)

\bibitem{STAR:2003vqj}
J.~Adams et~al. (STAR), Phys. Rev. Lett. \textbf{92}, 092301 (2004)

\bibitem{Shuryak:2002kd}
E.V. Shuryak, G.E. Brown, Nucl. Phys. A \textbf{717}, 322 (2003)

\bibitem{Kolb:2003bk}
P.F. Kolb, M.~Prakash, Phys. Rev. C \textbf{67}, 044902 (2003)

\bibitem{Rapp:2003ar}
R.~Rapp, Nucl. Phys. A \textbf{725}, 254 (2003)

\bibitem{Poskanzer:1998yz}
A.M. Poskanzer, S.A. Voloshin, Phys. Rev. C \textbf{58}, 1671 (1998)

\bibitem{STAR:2015gge}
L.~Adamczyk et~al. (STAR), Phys. Rev. Lett. \textbf{116}, 062301 (2016)

\bibitem{Dong:2004ve}
X.~Dong, S.~Esumi, P.~Sorensen, N.~Xu, Z.~Xu, Phys. Lett. B \textbf{597}, 328
  (2004)

\end{thebibliography}
\end{document}